\documentclass[11pt,twoside]{article}

\usepackage{asp2006}
\usepackage{epsf}
\usepackage{psfig}
\usepackage{lscape}
\usepackage{graphicx}

\begin{document}
\title{Origin and Nature of Dust in the Early Universe}   %%% Fill in title
\author{
Takaya Nozawa$^1$, 
Takashi Kozasa$^2$, 
Hideyuki Umeda$^3$,
Hiroyuki Hirashita$^4$,
Keiichi Maeda$^1$,
Ken'ichi Nomoto$^{1,3}$,
Nozomu Tominaga$^5$,
Asao Habe$^2$,
Eli Dwek$^6$,
Tsutomu T. Takeuchi$^7$,
Takako T. Ishii$^8$
}   %%% Fill in author names

\affil{
$^1$Institute for the Physics and Mathematics of the Universe, 
University of Tokyo, Kashiwa, Chiba 277-8568, Japan \\
$^2$Department of Cosmosciences, Graduate School of Science,
Hokkaido University, Sapporo 060-0810, Japan \\ 
$^3$Department of Astronomy, School of Science,
University of Tokyo, Bunkyo-ku, Tokyo 113-0033, Japan \\
$^4$Institute of Astronomy and Astrophysics, Academia Sinica,
Taipei 106, Taiwan \\
$^5$Division of Optical and Infrared Astronomy, National Astronomical 
Observatory of Japan, Mitaka, Tokyo 181-8588, Japan \\
$^6$Laboratory for Astronomy and Solar Physics,
NASA Goddard Space Flight Center, Greenbelt, MD 20771, USA \\
$^7$Institute for Advanced Research, Nagoya University, Nagoya 464-8601, 
Japan \\
$^8$Kwasan and Hida Observatories, Kyoto University, Yamashina, Kyoto
607-8471, Japan \\
}   %%% Fill in author affiliations

\begin{abstract} %%% Abstract to run on from here.

We present recent advances in theoretical studies of the formation and 
evolution of dust in primordial supernovae (SNe) that are considered to 
be the main sources of dust in the early universe.
Being combined with the results of calculations of dust formation in the 
ejecta of Population III SNe, the investigations of the evolution of newly 
formed dust within supernova remnants (SNRs) show that smaller grains are 
predominantly destroyed by sputtering in the shocked gas, while larger 
grains are injected into the ambient medium.
The mass of dust grains surviving the destruction in SNRs reaches up to
0.1--15 $M_\odot$, which is high enough to account for the content of 
dust observed for the host galaxies of quasars at $z > 5$.
In addition, the transport of dust formed in the ejecta causes the 
formation of low-mass stars in the dense shell of primordial SNRs and 
affects the elemental composition of those stars.
We also show that the flat extinction curve is expected in the 
high-redshift universe where SNe are the possible sources of dust.

\end{abstract}

\section{Introduction}   

Recent submillimeter and millimeter observations have detected the 
continuum thermal emission of dust from a growing number of quasars at 
redshifts larger than $z=5$ (Bertoldi et al.\ 2003; Priddey et al.\ 2003; 
Robson et al.\ 2004; Beelen et al.\ 2006; Wang et al.\ 2008a, 2008b).
The estimated mass of dust in the host galaxies of these quasars is 
10$^8$--10$^9$ $M_\odot$, which implies a rapid enrichment of the 
interstellar medium (ISM) with dust in such young systems.
In addition, the rest-frame near-infrared photometries of quasars at 
$z > 5$ with {\it Spitzer } (Hines et al.\ 2006; Jiang et al.\ 2006) have 
confirmed the existence of hot dust heated by the central active nuclei. 
This indicates that the fundamental structures of quasars such as 
accretion disks and hot-dust tori had been already established during the 
first billion years of cosmic history.
However, the origin and nature of dust in the high-redshift universe are 
not necessarily the same as those in our Galaxy.
In fact Maiolino et al.\ (2004a) have found that the rest-frame UV 
extinction curve observed for the quasar SDSS J1048+4637 at $z=6.2$ is 
quite different from those in local galaxies, suggesting that the source 
and evolution of dust at high redshifts ($z > 5$) are different from those 
at low redshifts ($z < 4$).

Dust grains in the early universe have great impacts on the formation 
processes of stars and galaxies.
Dust grains control the chemistry of the ISM by locking up a large 
fraction of refractory elements and by serving as catalysts for chemical 
reactions of gas.
Especially they foster the formation of hydrogen molecules on their 
surfaces (Cazaux \& Tielens 2004; Cazaux \& Spaans 2004), enhancing the 
star formation rate (SFR) in metal-poor star-forming gas clouds 
(Hirashita \& Ferrara 2002).
In dense metal-poor molecular clouds, dust grains provide efficient 
cooling pathways of gas through their thermal emission, which triggers 
the fragmentation of gas clouds into low-mass clumps of $\sim$0.1--1 
$M_\odot$ for metallicity of $10^{-6}$--$10^{-4}$ $Z_\odot$ (Schneider 
et al.\ 2002, 2003, 2006a; Omukai et al.\ 2005; Tsuribe \& Omukai 2006) 
and affects the initial mass function (IMF) in the metal-poor universe 
(e.g., Schneider et al.\ 2006b). 
Dust grains also influence the thermal balance in the interstellar 
space by producing photoelectrons that heat the gas and by reemitting
the absorbed stellar light at infrared wavelengths.

The reprocessing of stellar light by dust at high redshift is also
crucial in understanding the structure formation history of the universe 
from relevant observations.
Obscuration of stellar light by dust leads to a serious misleading in
interpreting the cosmic SFR and the IMF of the very early generation of 
stars from the cosmic infrared background radiation 
(e.g., Hauser \& Dwek 2001).
Thermal emission from high-$z$ dust dominates the far-infrared background
radiation and can even distort the cosmic microwave background radiation 
(Loeb \& Haiman 1997).
Thus, the investigation of dust in the early universe is one of the most 
important subjects to reveal the nature of stellar populations and 
underlying physical processes at the dark age of the universe.

In the previous studies that assessed the effects of dust grains at high 
redshift, the composition and size distribution of dust grains  
were assumed to be the same as those in our Galaxy, and their amount 
was treated as a parameter (e.g., Loeb \& Haiman 1997).
However, the physical processes involving the interactions of dust with 
photons and gas are sensitive to the chemical composition and size of dust.
The size distribution and amount of dust are determined by the competition 
between its production and destruction and vary with time according to 
the star formation activity.
Therefore, in order to evaluate the impacts of high-$z$ dust, it is 
essential to clarify 
the evolution of dust in the early epoch of the universe by taking account 
of the formation and destruction of dust self-consistently.

At redshift larger than 5 when the cosmic age is less than 1.2 billion 
years, the formation of dust in mass-loss winds from asymptotic giant 
branch (AGB) stars is too inefficient to supply a copious amount of dust 
into the ISM because most of low-mass stars cannot evolve off the main 
sequence in such an early epoch.
Hence, the main formation sites of dust are considered to be in the 
ejecta of Type II SNe (SNe II) evolving from short-lived massive stars 
with the progenitor masses $M_{\rm pr}=$ 8--40 $M_\odot$.
Indeed the chemical evolution models point out that 0.1--1 $M_\odot$ of 
dust per SN II event is required to form to explain a large content 
of dust observed in high-redshift galaxies
(Morgan \& Edmunds 2003; Maiolino et al.\ 2006; Dwek et al.\ 2007).
On the other hand, numerical simulations (e.g., Bromm et al.\ 2002;
Abel et al.\ 2002; Yoshida et al.\ 2006) and semi-analytic models 
(Nakamura \& Umemura 2001, 2002) have shown that the first stars formed 
out of the primordial composition of gas are typically very massive with 
masses more than 100 $M_\odot$.
Such metal-free stars as massive as 140--260 $M_\odot$ are considered to 
explode as pair-instability SNe (PISNe, Fryer et al.\ 2001; Umeda \& Nomoto 
2002; Heger et al.\ 2002) and pollute the ISM by ejecting large amounts of 
metal and dust at the end of their lives. 

Dust grains newly condensed in the SN ejecta are subsequently processed 
due to sputtering in the hot gas swept up by the reverse and forward 
shocks, and their size and mass can be greatly modified before being 
injected into the ISM.
Thus, in order to clarify what kind of and how much dust are supplied to 
the ISM by SNe, it is necessary to investigate the formation of dust in 
the ejecta and its evolution in the shocked gas within the SN 
remnants (SNRs).
In this proceedings, we describe the properties of dust in the early
universe revealed by a series of our studies;
the calculations of dust formation in the ejecta of primordial SNe II 
and PISNe are presented in \S~2, and the evolutions of dust in those SNRs 
are described in \S~3.
In \S~4 we discuss the role of dust in the early universe, focusing on 
the impacts of dust on the formation processes of the next generation 
of stars.
In \S~5 the extinction curve expected at high redshift is derived based
on the results of calculations given in \S~3 and is compared with the 
existing observational results.
The concluding remarks are presented in \S~6.

\section{Formation of dust in primordial SNe}

In the expanding ejecta of SNe, dust grains can condense through the 
formation of seed nuclei and their growth by the collisions with the gas 
in the metal-rich He core.
Based on the nucleation and grain growth theory by Kozasa \& Hasegawa 
(1987), Kozasa et al.\ (1989, 1991) demonstrated the formation sequence of 
dust in the ejecta of SN 1987A.
They also discussed that the composition of dust grains formed in the 
ejecta is controlled by the elemental composition of gas at their 
formation sites and that their radii and number densities strongly depend 
on the time evolution of gas temperature in the ejecta.

Applying the method of dust formation calculations by Kozasa et al.\ 
(1989, 1991), Todini \& Ferrara (2001) performed the calculations of 
dust formation in SNe II with the progenitor metallicity of $Z=$ 0--0.02.
They assumed one-zone models for the composition and density of gas 
within the He core, adopting the elemental yields for SNe II with
$M_{\rm pr} =$ 12--35 $M_\odot$ by Woosley \& Weaver (1995).
Considering the formation and destruction of CO and SiO molecules,
Todini \& Ferrara (2001) showed that C grains can condense along with 
silicate and oxide grains in the O-rich environment and that their size 
distributions are lognormal-like with the typical radii of 0.03 $\mu$m 
for C grains, 0.002 $\mu$m for Fe$_3$O$_4$ grains, and 0.001 $\mu$m for 
MgSiO$_3$ and Mg$_2$SiO$_4$ grains. 
The total mass of dust formed ranges from 0.08 $M_\odot$ to 2 $M_\odot$ 
depending on the progenitor mass and metallicity.
It should be noted that the one-zone uniform model for gas density and 
elemental composition with the same time evolution of gas temperature 
tuned by comparing with the observation of SN 1987A is not always 
applicable to clarify the dependence of $M_{\rm pr}$ and $Z$ on the 
total mass, chemical composition and size of dust formed in the ejecta.

\begin{figure}
\plottwo{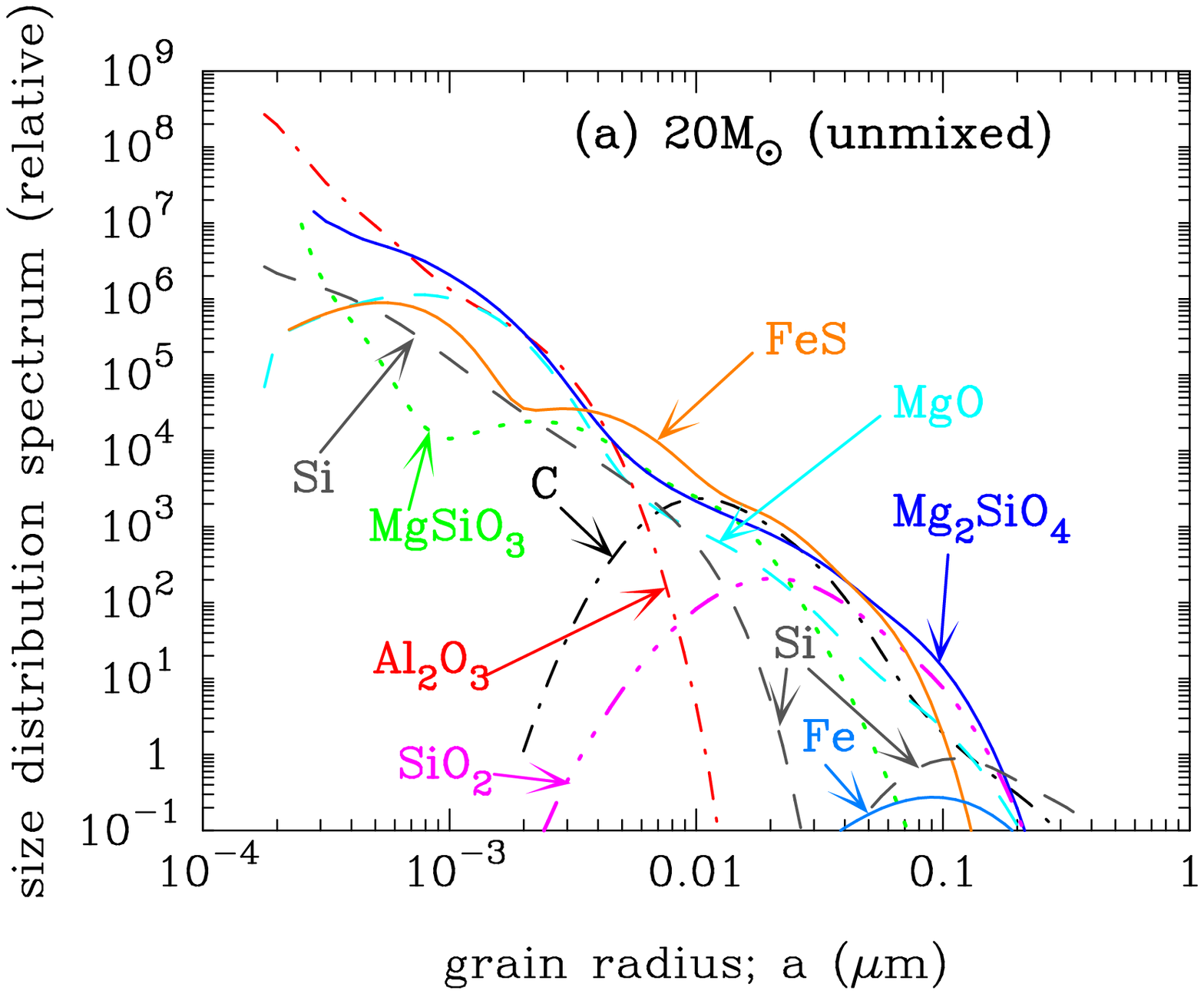}{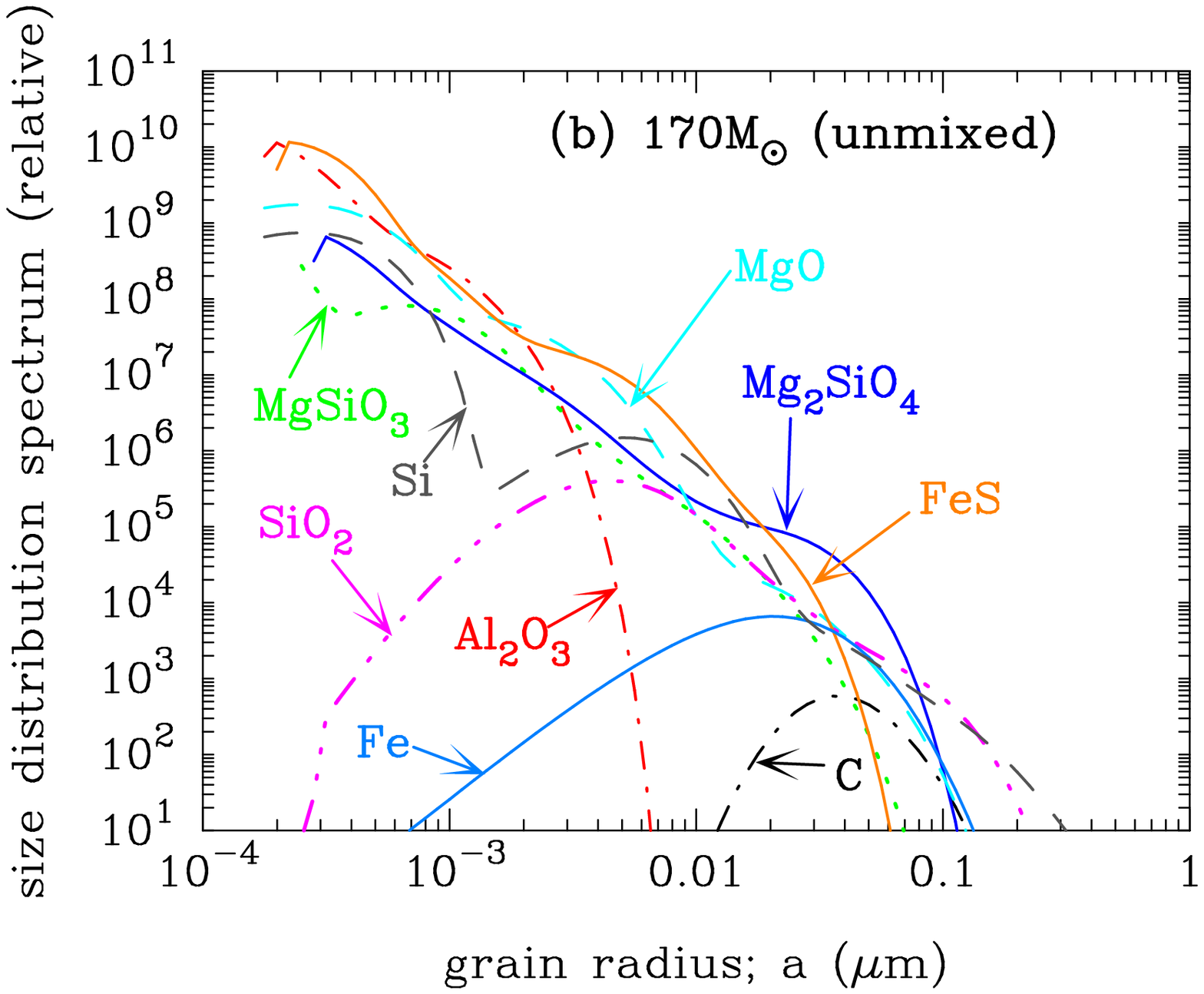}
\plottwo{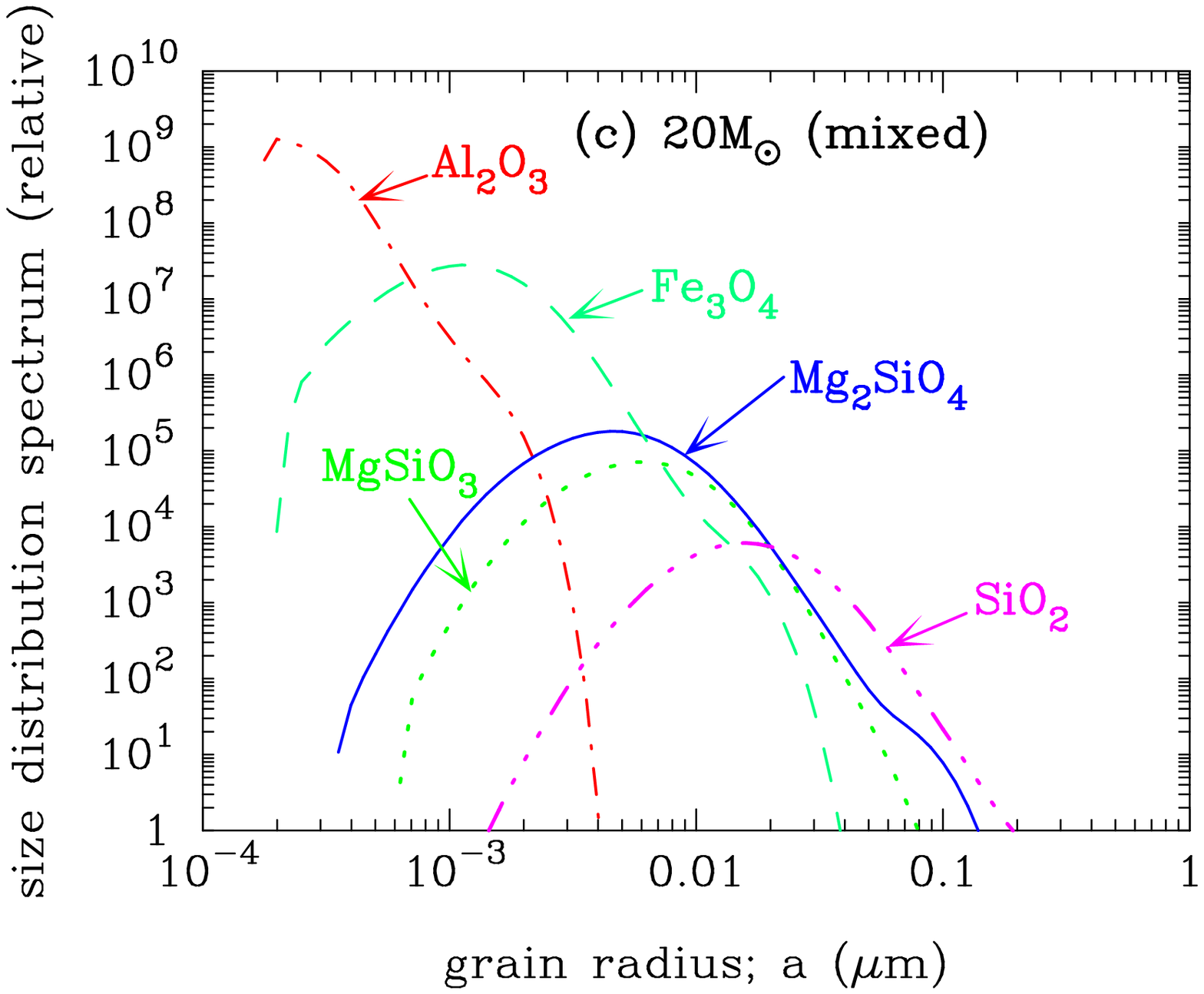}{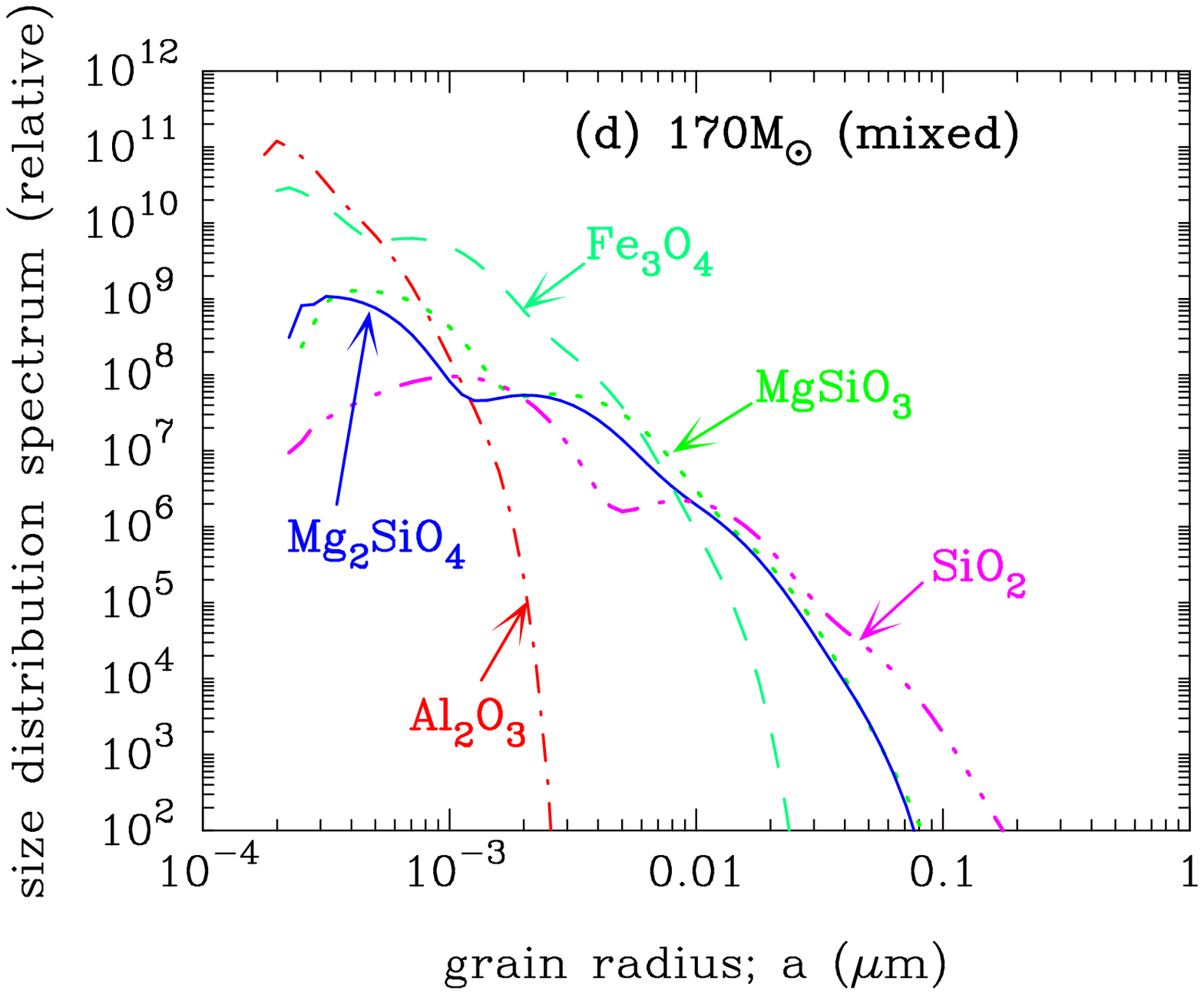}
\caption{
  Size distribution function of each grains species summed up over the 
  formation region in the unmixed ejecta
  (\textit{a}) for $M_{\rm pr}=$ 20 $M_\odot$ and
  (\textit{b}) for $M_{\rm pr}=$ 170 $M_\odot$, 
  and in the mixed ejecta 
  (\textit{c}) for $M_{\rm pr}=$ 20 $M_\odot$ and 
  (\textit{d}) for $M_{\rm pr}=$ 170 $M_\odot$.
}
\end{figure}

We investigated the condensation of dust in Population III SNe II and 
PISNe (Nozawa et al.\ 2003), adopting the hydrodynamic models and the 
results of nucleosynthesis calculations by Umeda \& Nomoto (2002).
We consider SNe II with $M_{\rm pr} =$ 13, 20, 25, and 30 $M_\odot$, 
all of which have the explosion energy of $10^{51}$ ergs, and PISNe with
$M_{\rm pr} =$ 170 $M_\odot$ and 200 $M_\odot$ whose explosion energies 
are $2 \times 10^{52}$ ergs and $2.8 \times 10^{52}$ ergs, respectively.
As for the mixing of elements within the He core, we consider the two 
extreme cases; the unmixed case with the onion-like elemental composition 
and the uniformly mixed case. 
The gas temperature is calculated for each model by solving the radiative 
transfer, taking account of the energy deposition from radioactive elements.
The formation of CO and SiO molecules is assumed to be complete.

Our results of dust formation calculations show that there is a great 
difference in the species of dust formed in the SN ejecta between the 
unmixed case and the mixed case. 
In the unmixed case, various grain species condense reflecting the 
difference of elemental composition in each layer: C grains in carbon-rich
He layer, Mg$_2$SiO$_4$, Al$_2$O$_3$, and MgO grains in O-Mg-Si layer,
MgSiO$_3$, Al$_2$O$_3$, and SiO$_2$ grains in O-Si-Mg layer, Si and FeS 
grains in Si-S-Fe layer, and Fe grains in the innermost Fe-rich layer.
In the mixed case, where the oxygen atoms are more abundant than carbon 
atoms, only silicates (MgSiO$_3$, Mg$_2$SiO$_4$, SiO$_2$) and oxide 
(Al$_2$O$_3$ and Fe$_3$O$_4$) grains form. 

Figure 1 shows the differential size distribution of each dust species 
summed up over its formation region in the unmixed and mixed cases for 
SNe II (Figs.\ 1{\it a} and 1{\it c}) and PISNe (Figs.\ 1{\it b} and 
1{\it d}). 
In the unmixed case, the size distributions of C, SiO$_2$, and Fe grains 
are lognormal-like with relatively large radii ($\ga 0.01$ $\mu$m), while 
the other grain species have power-law-like distributions with radii 
ranging a few orders of magnitude.
In the mixed ejecta, the size distribution function of dust formed is 
almost lognormal-like except for Al$_2$O$_3$ grains. 
The average radius of dust for the mixed case is a little smaller compared 
with that for the unmixed case, although the maximum radius of dust is 
limited to less than 1 $\mu$m in both cases.
The main species of newly formed dust and their size distribution 
functions are almost independent of the progenitor mass.

\begin{figure}
\plottwo{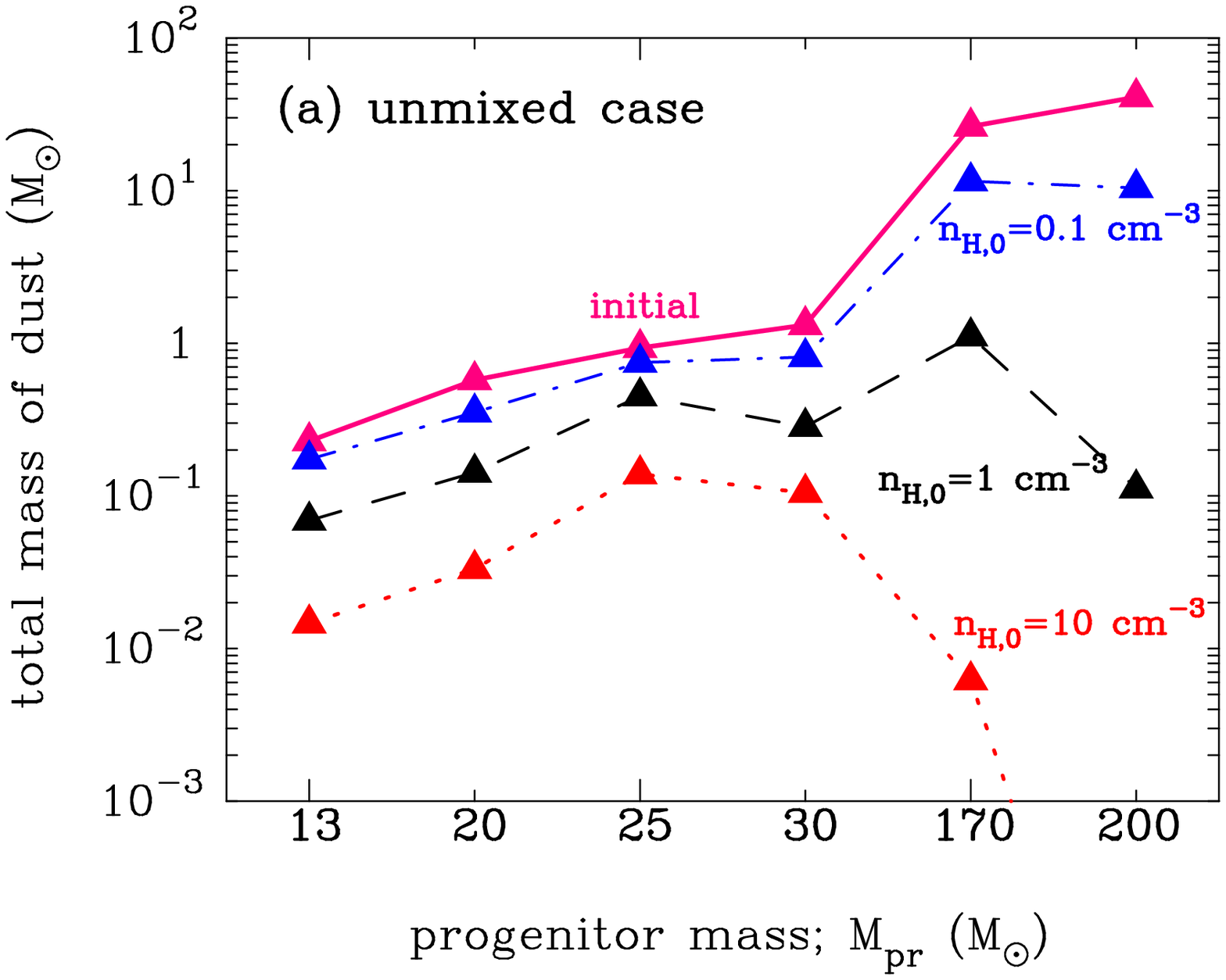}{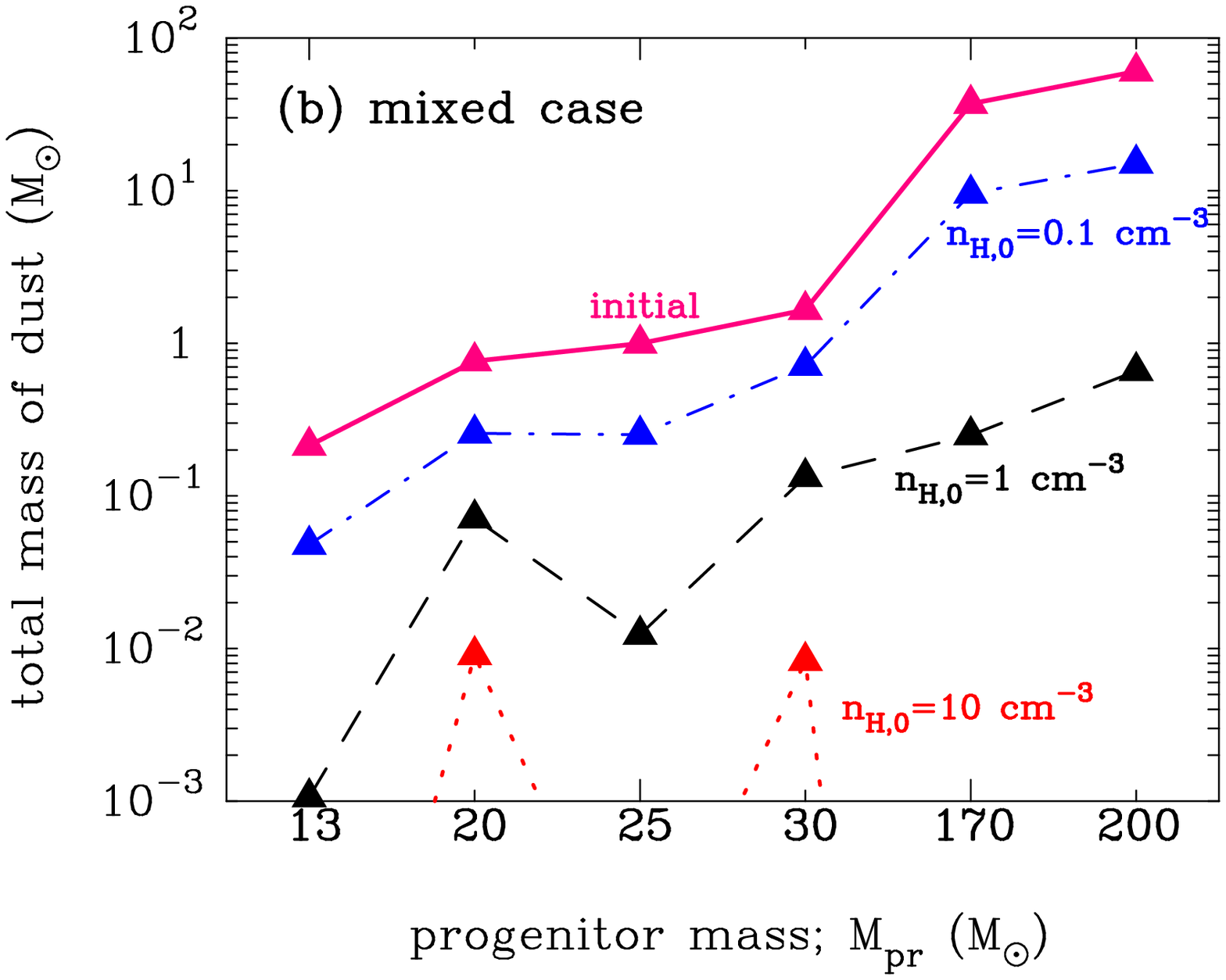}
\caption{
  Total mass of dust at the time of dust formation versus progenitor mass 
  ({\it connected by the solid lines}), together with total mass of dust 
  surviving the destruction in SNRs for various ambient gas densities
  (see \S~3);
  ({\it a}) for the unmixed case and ({\it b}) for the mixed case.
  The results for $n_{{\rm H},0} = 0.1$, 1, and 10 cm$^{-3}$ are connected
  by the dot-dashed, dashed, and dotted lines, respectively.
}
\end{figure}

The total mass of dust formed in the ejecta of primordial SNe is shown 
in Figure 2 for the unmixed case (Fig.\ 2{\it a}) and the mixed case 
(Fig.\ 2{\it b}). 
In general, the total mass of dust is higher for the mixed case than the 
unmixed case because oxygen atoms are more efficiently locked into dust 
grains as silicates and oxides in the mixed case.
The total mass of dust formed is 0.1--2 $M_\odot$ for Type II SNe and 
25--60 $M_\odot$ for PISNe and increases with increasing the progenitor 
mass.
The depletion factor, defined as the mass fraction of metal locked into 
grains to metal available in the ejecta, is 0.05--0.25 for SNe II
and 0.45--0.55 for PISNe.
Therefore, PISNe can convert a much larger fraction of refractory elements 
into dust grains than SNe II.
Note that the yields of dust in PISNe are consistent with the results by 
Schneider et al.\ (2004) who extended the model of dust formation in SNe II
by Todini \& Ferrara (2001) to PISNe, adopting the nucleosynthesis results 
by Heger \& Woosley (2002).

\section{Evolution of newly formed dust in primordial SNRs}   

Before newly formed dust grains in the ejecta are injected into the ISM, 
they can be processed in the hot gas swept up by the reverse and forward 
shocks that are generated by the interaction between the SN ejecta and 
the surrounding medium.
Dust grains condensed in the He core initially expand with the gas, 
but once they encounter the reverse shock, they acquire high velocities 
relative to the gas and are eroded by sputtering in the shocked gas.
Dust grains are also decelerated by the gas drag, and some of them are 
trapped in the hot plasma and are completely destroyed by sputtering.
The deceleration rate and erosion rate of dust depend on the temperature 
and density of gas and the relative velocity of dust to gas, as well as
the composition and size of dust.
Therefore, in order to reveal the composition, size distribution, and 
amount of dust supplied from SNe into the ISM, we need to investigate the 
motion and processing of dust within SNRs, based on the physical
properties of dust obtained by the dust formation calculations.

Bianchi \& Schneider (2007) have semi-analytically investigated the 
destruction of newly formed dust through the passage of the reverse 
shock, by revisiting the models of dust formation by Todini \& Ferrara 
(2001) and by employing a little different size distribution of dust 
from their results. 
Bianchi \& Schneider (2007) found that the size distribution of dust 
surviving the reverse shock is skewed to smaller radius in comparison 
with the original one.
The mass of surviving dust is 2--20 \% of the initial dust mass and 
decreases with increasing the gas density in the ambient medium.
However, in the calculations, they did not consider the motion of dust 
outwardly penetrating through the shocked hot gas, which is a very 
important process to pursue the evolution and survival of dust within 
SNRs.
Their calculations are preformed only up to $\la$ 10$^5$ years.

We investigated the evolution of dust in primordial SNRs (Nozawa et al.\ 
2007), by adopting the results of dust formation calculations by 
Nozawa et al.\ (2003) and by carefully treating the dynamics of dust and 
its destruction by sputtering.
Here we briefly introduce the method of calculation;
the deceleration rate of dust due to the gas drag is inversely proportional 
to the grain radius and bulk mass density.
The erosion rate by sputtering is evaluated by applying the sputtering 
yield for each dust species given in Nozawa et al.\ (2006).
Dust grains being treated as test particles, the transport and
destruction of dust are calculated in a self-consistent manner by taking 
account of the size distribution and spatial distribution of each grain 
species formed in the ejecta.

The time evolution of gas temperature and density in SNRs is numerically 
solved by assuming spherical symmetry and considering the cooling of gas 
via the atomic process, the inverse Compton, and the thermal radiation of 
dust.
For the initial conditions of gas in the ejecta, we adopt the hydrodynamic 
models of Population III SNe II and PISNe by Umeda \& Nomoto (2002), which 
are the same as those used in the dust formation calculations.
We assume the uniform ambient medium with a gas temperature of $10^4$ K 
based on the studies of the radiative feedback from massive primordial 
stars (Kitayama et al.\ 2004; Machida et al.\ 2005).
To examine the dependence of the evolution of dust on the gas density in 
the ISM, we consider the hydrogen number densities of $n_{\rm H,0}=$ 
0.1, 1, and 10 cm$^{-3}$ for the ambient gas density.
The calculations are carried out from 10 years up to 10$^6$ years.

\begin{figure}
  \includegraphics[scale=0.44, angle=0, viewport= -150 0 100 500]{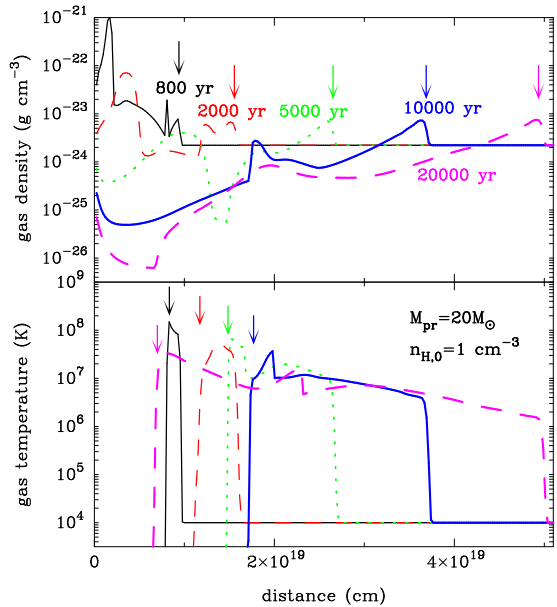}
\caption{
  Structures of gas density (\textit{upper panel}) and temperature 
  (\textit{lower panel}) at given times in the SNR that is generated 
  from the explosion of the star with $M_{\rm pr}=20$ $M_{\odot}$ and is 
  expanding into the ISM with $n_{\rm H,0}=1$ cm$^{-3}$.
  The downward-pointing arrows in the upper and lower panels indicate the 
  positions of the forward and reverse shocks, respectively.
}
\end{figure}

Figure 3 shows the time evolutions of the structure of gas density 
({\it upper panel}) and temperature ({\it lower panel}) in the SNRs for 
$M_{\rm pr}=$ 20 $M_\odot$ and ambient density of $n_{\rm H,0}=1$ 
cm$^{-3}$.
The positions of the forward and reverse shocks are indicated by the arrows 
in the upper and lower panel, respectively, at given times after explosion.
We can see that the gas swept up by the forward and reverse shocks is 
heated to the temperature of $10^6$--$10^8$ K, where dust grains are 
efficiently eroded by sputtering (see Fig.\ 2 in Nozawa et al.\ 2006).

\begin{figure}
  \includegraphics[scale=0.44, angle=0, viewport= -150 0 100 500]{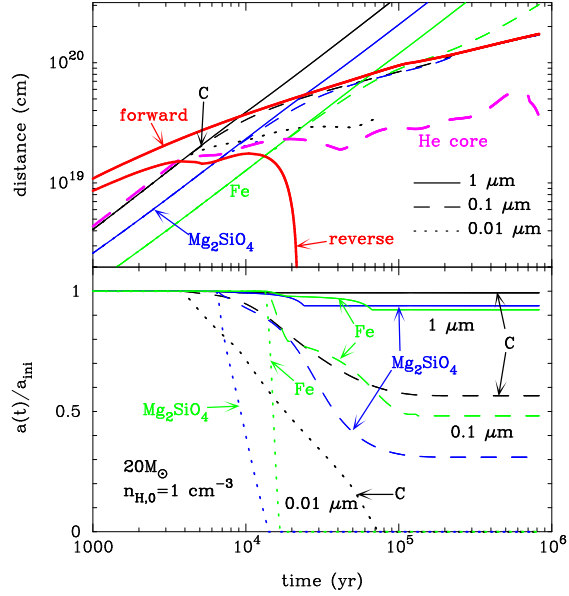}
\caption{
  Trajectories of C, Mg$_2$SiO$_4$, and Fe grains (\textit{upper panel}) 
  and the time evolutions of their radii relative to the initial ones
  (\textit{lower panel}) within the SNR given in Figure 3. 
  The evolutions of dust with $a_{\rm ini}=$ 0.01, 0.1, and 1 $\mu$m 
  are depicted by the dotted, dashed, and solid lines, respectively.
  In the upper panel, the trajectories of the reverse and forward shocks 
  and the outer boundary of He core are indicated by the thick solid and 
  dashed lines.
}
\end{figure}

Figure 4 exhibits the trajectory of dust ({\it upper panel}) and the time 
evolution of dust radius relative to the initial one ({\it lower panel}) 
within the SNR given in Figure 3.
The thick solid curves in the upper panel depict the trajectories of the 
reverse and forward shocks.
The evolution of dust within the SNR is shown for C, Mg$_2$SiO$_4$, and 
Fe grains formed in the unmixed case.
Initially dust grains are expanding with the gas, and then collide with 
the reverse shock at different times depending on their initial positions:
at 3650 yr for C grains formed in the outermost He core, at 6300 yr for
Mg$_2$SiO$_4$ grains in the O-rich layer, and at 13,000 yr for Fe grains
condensed in the innermost He core. 

Once dust grains intrude into the reverse shock, the evolution of dust 
heavily depends on the composition and initial radius $a_{\rm ini}$ 
(see Fig.\ 4).
Grains with the initial radius of $a_{\rm ini}=$ 0.01 $\mu$m 
({\it dotted lines}) are efficiently decelerated by the gas drag and are 
trapped in the hot gas between the reverse and forward shocks.
These small grains are completely destroyed by sputtering in the shocked 
gas.
Grains of $a_{\rm ini}=$ 0.1 $\mu$m ({\it dashed lines}) reduce their sizes 
by sputtering but are finally trapped in the dense shell formed behind the 
forward shock at $2 \times 10^5$ years without being completely destroyed.
Fe grains of $a_{\rm ini}=$ 0.1 $\mu$m are injected into the ambient medium 
because of their high bulk density.
Grains with $a_{\rm ini}=$ 1 $\mu$m ({\it solid lines}) can maintain their 
high velocities and pass through the forward shock to get injected into 
the ambient medium without being eroded significantly.

In Figure 5, we show the size distribution function of each dust species
surviving the destruction in SNRs.
Figure 5{\it a} presents the size distribution of dust in the unmixed case 
after destruction for $M_{\rm pr}=20$ $M_\odot$ and $n_{\rm H,0}=1$ 
cm$^{-3}$, and Figure 5{\it b} for $M_{\rm pr}=170$ $M_\odot$ and 
$n_{\rm H,0}=0.1$ cm$^{-3}$.
Because small grains are efficiently trapped in the hot gas and are 
predominantly destroyed by sputtering, the size distribution of surviving 
dust is greatly deficient in small-sized grains, compared with that at its 
formation for both cases of SNe II and PISNe 
(see Figs.\ 1{\it a} and 1{\it b} for their initial size distributions).
Thus, we conclude that relatively large-sized grains inhabit the 
interstellar space in the early universe, where dust grains are supplied 
only by SNe II and PISNe.

\begin{figure}
\plottwo{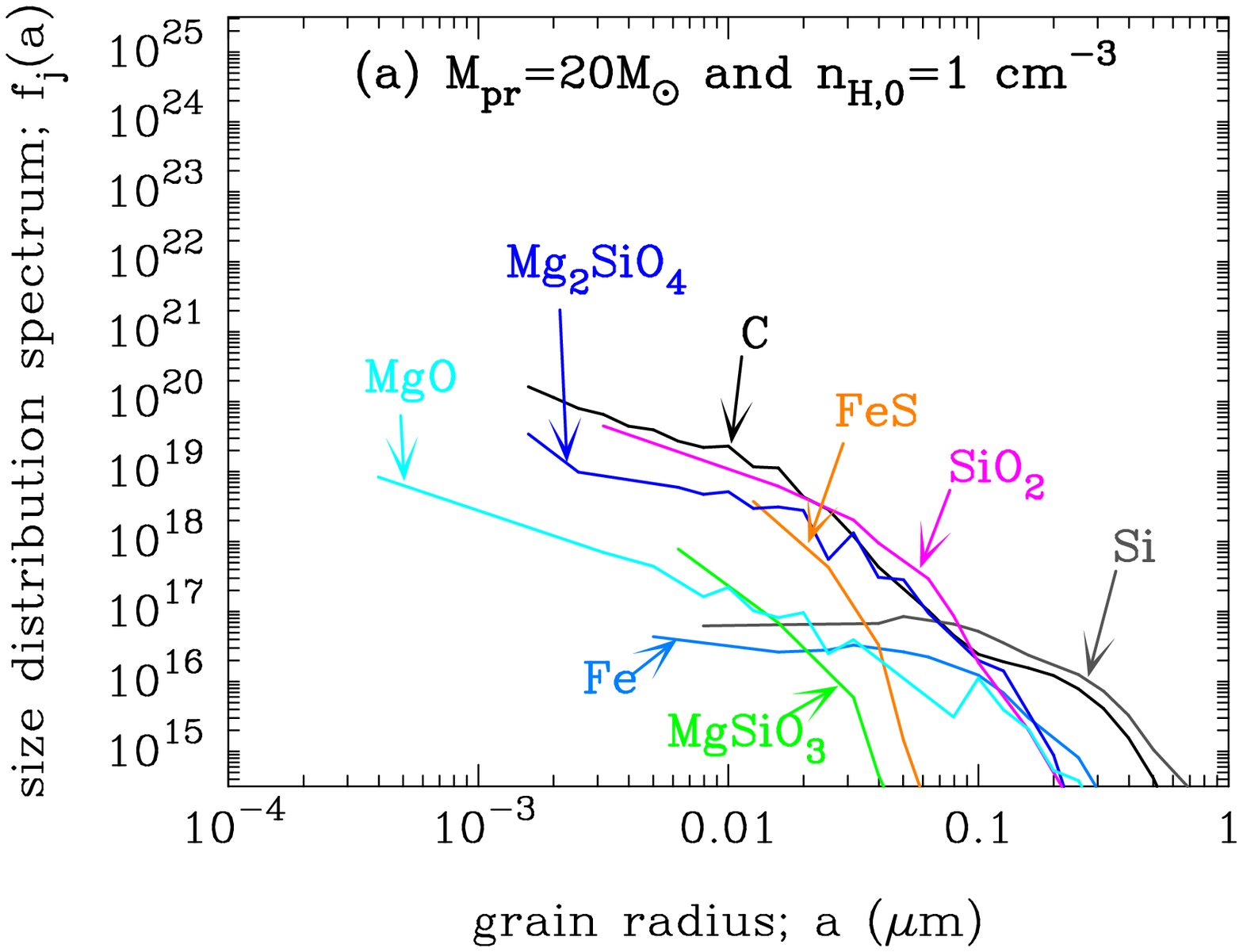}{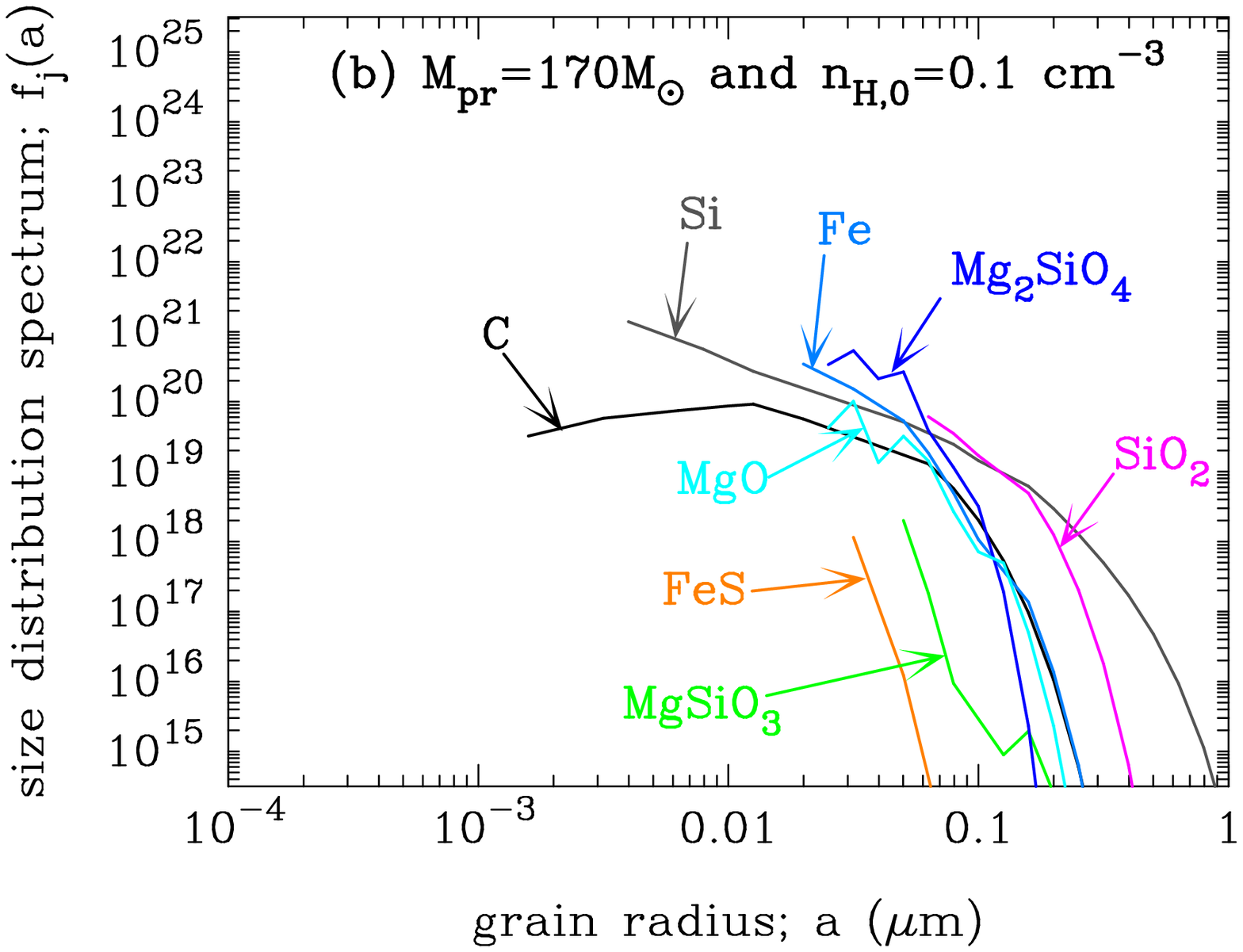}
\caption{
  Size distribution of each dust species surviving the destruction in SNRs
  for the unmixed case with ({\it a}) $M_{\rm pr}=20$ $M_{\odot}$ and 
  $n_{\rm H,0}=1$ cm$^{-3}$ and ({\it b}) $M_{\rm pr}=170$ $M_{\odot}$ and 
  $n_{\rm H,0}=0.1$ cm$^{-3}$.
}
\end{figure}

The behaviors of size distribution of surviving dust do not depend on the 
progenitor mass for SNe II because the explosion energy is the same 
($E_{51}=1$) and thus the time evolution of temperature and density of gas 
in a SNR is similar.
For $n_{\rm H,0}=1$ cm$^{-3}$, the critical radius below which dust is 
completely destroyed is $\sim$0.05 $\mu$m although it weakly depends on 
the dust species and their initial positions.
For PISNe, the critical radius is approximately three times higher than 
that for SNe II as the result of much higher explosion energy
($>$$10^{52}$ ergs) and more massive hydrogen envelope compared to SNe II.

The evolution of dust in SNRs is also affected by the ambient gas density.
A higher ambient density results in denser shocked gas, which causes the 
efficient erosion and deceleration of dust due to more frequent collisions 
with the hot gas.
Hence, the critical radius for each dust species increases with increasing 
ambient gas density: $\sim$0.01 $\mu$m for $n_{\rm H,0}=0.1$ cm$^{-3}$ 
and $\sim$0.2 $\mu$m for $n_{\rm H,0}=10$ cm$^{-3}$.
Accordingly, the total mass of surviving dust decreases with increasing 
gas density in the ISM, as shown in Figure 2.
For $n_{\rm H,0}=10$ cm$^{-3}$, the mass of surviving dust is less than
0.1 $M_\odot$ for all models considered here, and all or almost all 
($\ga$ 85 \%) of the dust grains formed in the ejecta are destroyed.
The mass fraction of dust destroyed is generally larger for the mixed case
than the unmixed case because the mass of dust formed in the mixed case is 
somewhat weighted toward smaller size than that in the unmixed case. 
It can be also seen from Figure 2 that the mass fraction of dust destroyed
in remnants of PISNe is much larger than in SNe II.
This is because the critical radius is large and the average radius of dust
formed is relatively small for PISNe.
Nevertheless, for $n_{\rm H,0}=0.1$ and 1 cm$^{-3}$, the mass of dust 
surviving the destruction is between 0.1 $M_\odot$ and 15 $M_\odot$, which 
is high enough to account for the dust budget observed in the host 
galaxies of quasars at $z > 5$ (Maiolino et al.\ 2006; Dwek et al.\ 2007).
Thus, we conclude that SNe II and PISNe are the important dust factories 
in the early universe.

It should be mentioned here that although we have focused on Population III 
SNe, the results of the present study could be useful for evaluating the 
composition, size distribution, and mass of dust ejected from SNe II with 
high metallicity of the progenitors.
One of the reasons for this is that the species of dust formed in the 
ejecta of SNe II and their size distribution are insensitive to the 
metallicity of the progenitor stars (Todini \& Ferrara 2001; Nozawa 2003).
On the other hand, the increase of the metallicity in the ambient medium 
enhances the cooling in the shocked gas and affects the efficiency of dust 
destruction in a SNR.
However, even if we calculate the evolution of dust within a SNR for the 
ambient medium with solar metallicity, the destruction efficiency of each 
grain species decreases at most by 15 \% compared with that for the 
primordial gas composition (Nozawa et al.\ 2007).
Therefore, the results of calculations described here can be also
applicable to Type II-p SNe in the local universe, regardless of the 
initial metallicity of progenitor stars and ambient medium.

\section{Role of dust on the formation process of stars}   

In the previous sections, we have shown that dust grains can condense even 
in the ejecta of Population III SNe II and PISNe evolving from zero-metal 
progenitor stars (\S~2) and that a sufficient fraction of newly formed 
grains can survive the destruction in the SNRs as long as the gas density 
in the surrounding medium is not very high (\S~3).
This indicates that dust grains are surely ejected from primordial SNe and 
can quickly enrich the ISM in the early universe.
In this section we describe the fundamental role that the first cosmic 
dust plays in the formation processes of the subsequent generations of 
stars.

\subsection{Critical metallicity}   

The theoretical studies investigating the collapse and fragmentation of 
the primordial gas (e.g., Bromm \& Larson 2004 and references therein) 
have predicted that the characteristic mass of the first stars is very 
high, probably in excess of 100 $M_\odot$. 
On the other hand, a typical mass of Population I and II stars seen in 
the present days is as low as $\sim$1 $M_\odot$. 
Thus, in the course of cosmic history, there should have been the 
transition of star formation mode from massive Population III stars to 
low-mass Population II stars.

The fundamental quantity responsible for this transition is the initial
metallicity of collapsing gas clouds (Yoshii \& Sabano 1980; Omukai 2000).
The presence of metals largely enhances the cooling efficiency of gas
and governs the thermal and chemical evolution of low-metallicity clouds.
If the initial metallicity reaches a critical value $Z_{\rm cr}$, 
the protostellar clouds become gravitationally unstable at higher 
densities where the Jeans masses are correspondingly lower, and undergo 
fragmentation to produce low-mass gas clumps.
Bromm et al.\ (2001) computed the evolution of pre-enriched star-forming
gas clouds and found that the critical metallicity is likely to be
$Z_{\rm cr} \simeq 5 \times 10^{-4}$ $Z_\odot$.
Also, Bromm \& Loeb (2003) and Santro \& Shull (2006) have 
derived the critical abundance of each heavy element such as C, O, Si, and 
Fe, above which the fragmentation of gas clouds is caused by
fine-structure line cooling of the individual heavy element.

It should be noted here that if a significant fraction of metals are 
locked into dust grains, cooling by thermal emission of dust is more 
efficient than metal line cooling in the dense metal-poor molecular 
clouds.
Schneider et al.\ (2006a) investigated the thermal history of collapsing 
gas clouds, developing the calculations by Omukai et al.\ (2005) and 
adopting the models of dust formed in primordial SNe II by Todini \& 
Ferrara (2001) and PISNe by Schneider et al.\ (2004).
They concluded that the efficient cooling of gas by dust grains enables 
fragmentation of gas clouds into solar or sub-solar mass scales even at 
$Z_{\rm cr} = 10^{-6}$ $Z_\odot$.
Their calculations are based on one-zone models for the evolution of 
clouds, but this prediction is confirmed by the 3-D hydrodynamical 
simulations. 
Tsuribe \& Omukai (2006) studied the dynamical collapse and fragmentation 
of gas clouds with $Z = 10^{-6}$--10$^{-5}$ $Z_\odot$ and showed that the 
core of prestellar clouds actually fragments into small gas clumps.
Therefore, dust produced in primordial SNe plays a critical role in the 
formation of the first low-mass stars in the universe.

\subsection{Impact of dust on Population II.5 stars}   

In the halo of our Galaxy, many extremely metal-poor (EMP) low-mass 
stars with [Fe/H] $< -3$ have been discovered so far. 
In particular, two hyper-metal-poor (HMP) stars, HE 0107--5240 
(Christlieb et al.\ 2002) and HE 1327--2326 (Frebel et al.\ 2005) 
with [Fe/H] $< -5$, and an ultra-metal-poor (UMP) star, HE 0557--4840 
(Norris et al.\ 2007) with [Fe/H] $< -4$, are considered to be very early 
generation of stars.
The most iron-deficient HMP stars show extreme ($\ge$100 times)
overabundances of C and O and modest (1--100 times) enhancements of Si 
and Mg, and several scenarios have been proposed to explain the origin 
of their peculiar elemental compositions
(Beers \& Christlieb 2005 and references therein). 
Here we would suggest one possibility that dust grains newly formed in 
primordial SNe are responsible for the elemental compositions of these 
low-mass stars, considering that they are the second generation of stars 
formed in the dense shell of primordial SNRs, which are referred to as 
Population II.5 stars
(Mackey et al.\ 2003; Salvattera et al.\ 2004; Machida et al.\ 2005).

As demonstrated in \S~3, the results for the evolution of dust within a 
SNR show that the dust grains surviving the destruction but not injected 
into the ambient medium are accumulated in the dense shell behind the 
forward shock after 10$^5$ to 10$^6$ years.
The existence of dust in the shell may make possible the formation of stars 
with solar mass scales through its thermal emission if its metallicity 
is above 10$^{-6}$ $Z_\odot$, as shown in the last subsection.
In that case, it is expected that the elemental composition of piled-up 
grains reflects the elemental abundance of low-mass Population II.5 stars.
Thus, we analyze the metallicity and metal abundance in the SN shell and 
compare with observations of HMP and UMP stars.

\begin{table}
\begin{center}
  \caption{
Metallicities, [Fe/H], and abundances of C, O, Mg, and Si relative to Fe 
in the dense shell of primordial SNRs for various progenitor masses and 
ambient gas densities. 
The results calculated for the unmixed case of SNe II are only presented 
(see text).}
  \label{tab1}

\vspace{0.2cm}

\begin{tabular}{lcccccc}

\hline
  {$M_{\rm pr}$} & 
{$\log(Z/Z_\odot)$} & ~{[Fe/H]}~ & 
~{[C/Fe]}~ & ~{[O/Fe]}~ & ~{[Mg/Fe]}~ & ~{[Si/Fe]}  \\
\hline
\multicolumn{7}{c}{$n_{\rm H,0}=0.1$ cm$^{-3}$} \\
\hline
13 & $-5.89$ & $-6.43$ & $-0.274$ & $-0.699$ & $-0.230$ & $1.92$ \\
20 & $-5.44$ & $-5.20$ & $0.117$  & $-0.595$ & $0.034$  & $0.410$ \\
25 & $-5.55$ & $-5.90$ & $1.11$   & $-1.42$  & $-0.500$ & $-0.552$ \\
30 & $-5.33$ & $-5.56$ & $0.566$  & $-0.043$ & $0.739$  & $0.866$ \\
\hline
\hline
\multicolumn{7}{c}{$n_{\rm H,0}=1$ cm$^{-3}$} \\
\hline
13 & $-4.72$ & $-5.15$ & $1.11$  & $-0.555$ & $-0.459$ & $1.01$ \\
20 & $-4.68$ & $-5.53$ & $0.992$ & $0.585$  & $1.16$   & $1.87$ \\
25 & $-4.79$ & $-5.23$ & $1.09$  & $-0.412$ & $0.407$  & $0.989$ \\
30 & $-4.60$ & $-5.11$ & $0.797$ & $0.242$  & $1.09$   & $1.26$  \\
\hline
\hline
\multicolumn{7}{c}{$n_{\rm H,0}=10$ cm$^{-3}$} \\
\hline
13 & $-4.40$ & $-4.13$ & $0.284$  & $-2.54$ & $-3.89$ & $0.599$ \\
20 & $-4.09$ & $-4.92$ & $0.946$  & $-2.15$ & $-1.80$ & $2.14$ \\
25 & $-3.91$ & $-5.10$ & $1.60$   & $0.122$ & $0.232$ & $2.34$ \\
30 & $-3.84$ & $-5.11$ & $-0.207$ & $0.375$ & $-1.23$ & $2.66$ \\
\end{tabular}
\end{center}
\end{table}

Table 1 presents the elemental composition and metallicity in the dense 
SN shell for various progenitor masses and gas densities in the ambient 
medium.
We show only the results calculated for the unmixed case of SNe 
II since Fe-bearing grains are not significantly piled up in the shell 
of SNRs for the mixed case of SNe II and both cases of PISNe.
Table 1 shows that metallicity of the gas in the shell is in the range 
from 10$^{-6}$ to 10$^{-4}$ $Z_\odot$, which satisfies the condition to 
cause the formation of stars with solar mass scales 
(Schneider et al.\ 2006a; Tsuribe \& Omukai 2006).
We can also see that most of calculated [Fe/H] spans the range from $-6$ 
to $-4.5$ and are in good agreement with those for HMP and UMP stars.
Thus, we can conclude that the transport of dust segregated from 
metal-rich gas can be an important process to form the iron-poor low-mass
stars such as HMP and UMP stars.

In Figure 6{\it a}, we also plot the abundances of C, O, Mg, and Si 
in the dense shell relative to solar value for the ambient densities of 
0.1 and 1 cm$^{-3}$.
The results are denoted by triangles, while the observational data of HMP 
and UMP stars are plotted by other symbols.
The calculated abundances of Mg and Si also reproduce 1--100 times 
overabundances in HMP stars.
Therefore, the elemental composition of dust piled up in the shell can 
influence the abundance patterns of metals such as Fe, Mg, and Si in HMP 
and UMP stars if they are Population II.5 stars.
However, more than 100 times excesses of C and O cannot be reproduced by 
any models considered here.
For one of the reasons, we assumed that the metal-rich gas in the ejecta 
does not mix with the gas in the shell.

\begin{figure}
\plottwo{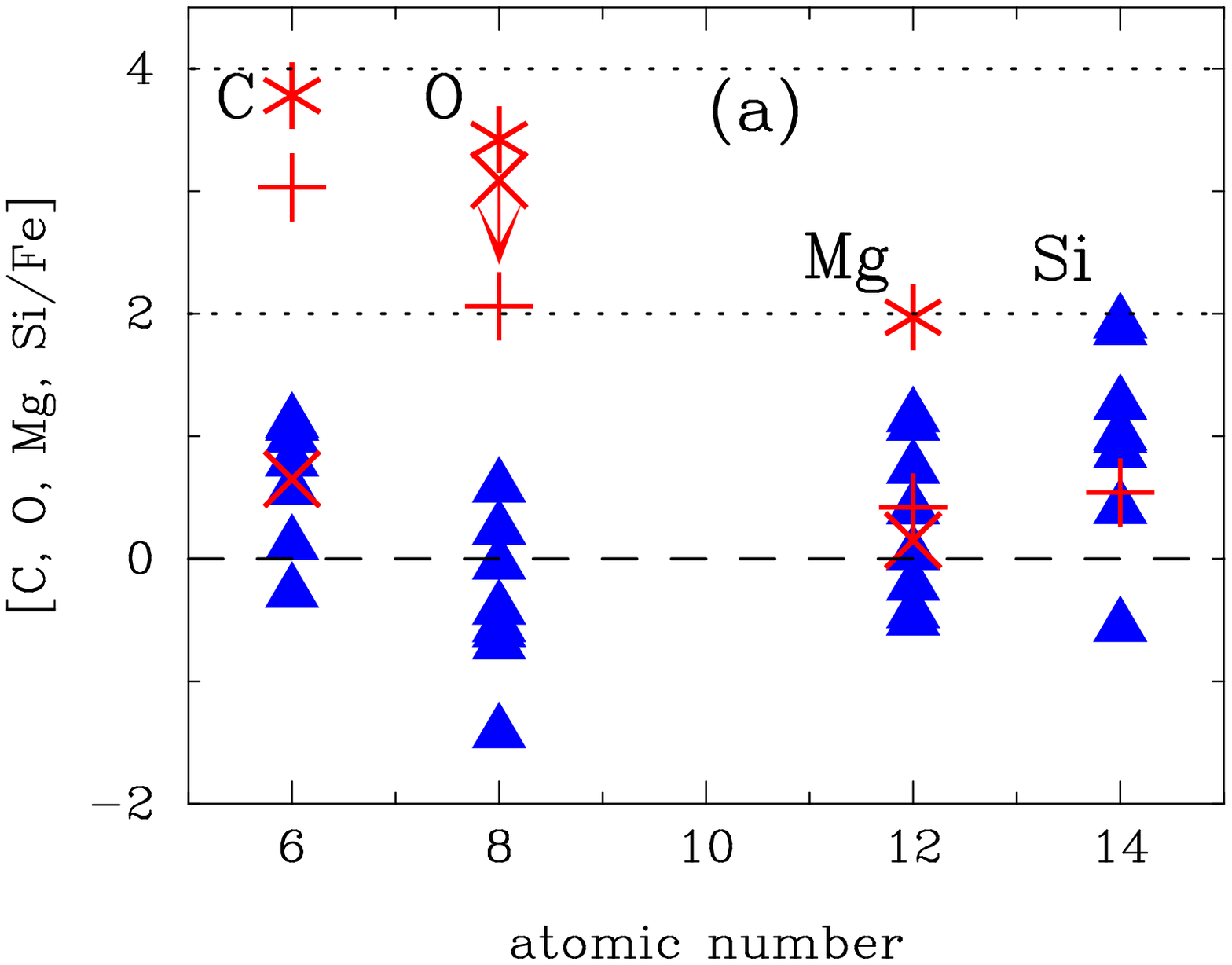}{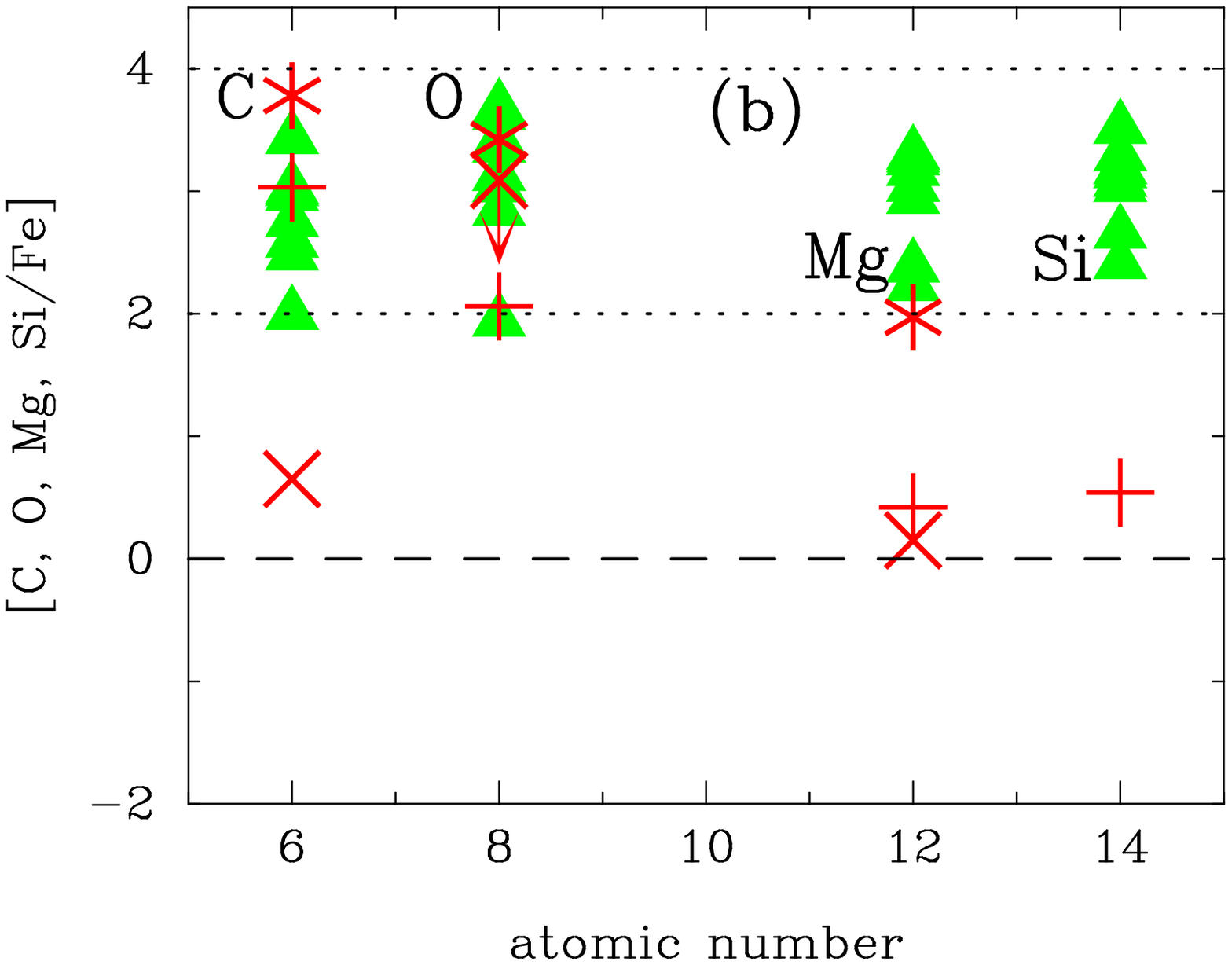}
\caption{
  Abundances of C, O, Mg and Si relative to solar value in the dense shell 
  of primordial SN II remnants for $n_{\rm H,0} = 0.1$ and 1 cm$^{-3}$ 
  ({\it filled triangles}); 
  (\textit{a}) derived from the elemental composition of the 
  grains piled up in the shell, and (\textit{b}) derived from the elemental 
  composition of the piled-up grains and the gas outside the innermost 
  Fe layer.
  For the observational data of HMP and UMP stars, the 3-D corrected 
  abundances are adopted and are denoted by plus (HE0107--5240 with 
  [Fe/H] = $-5.62$, Collet et al.\ 2006), asterisk (HE1327--2326 with 
  [Fe/H] = $-5.96$, Frebel et al.\ 2008), and cross (HE0557--4840 with
  [Fe/H] = $-4.75$, Norris et al.\ 2007). }
\end{figure}

Hence, we then examine the abundance patterns in the shell by considering 
that in addition to the piled-up grains, the gas outside the innermost 
iron layer of the SNe is incorporated into the shell until the time of star 
formation.
The results are shown in Figure 6{\it b}.
This case can produce the extreme overabundances of C and O, but leads to 
unreasonable excesses ($\ge$ 100 times) of Mg and Si as well, which is not 
in agreement with the observations.
It might be possible to reproduce the elemental abundances of 
HMP stars if the Si-Mg-rich layer is not mixed with the gas in the shell,
although it is necessary to clarify what extent of the gas in the ejecta 
can be actually incorporated into the shell 
when Population II.5 stars form. 

We also mention here that Venkatesan et al.\ (2006) have proposed that 
the peculiar abundances of refractory elements in EMP stars can be 
reproduced by the transport of newly formed dust decoupled from the 
metal-rich gas in primordial SNe II and PISNe with $M_{\rm pr}=$
10--150 $M_\odot$.
However, they considered that the transport of dust is driven by the UV 
radiation field from the stellar cluster within a SNR, which seems to be 
an impracticable assumption.
Furthermore, they have not made a detailed comparison between their 
results and the abundance data of EMP stars.

\section{Extinction curve expected at the high-redshift universe}   

The properties of dust at high redshift have been probed by the dust 
extinction curves for quasars at $z > 4.9$ (Maiolino et al.\ 2004a).
In particular, broad absorption line (BAL) quasars are good targets on this 
matter because they are associated with powerful outflows of dense gas and 
are significantly reddened by dust. 
It has been also known that the extinction properties for low-$z$ 
($z < 4$) BAL quasars are well described by the extinction curve in the 
Small Magellanic Cloud (SMC)
(Reichard et al.\ 2003; Richard et al.\ 2003; Hopkins et al.\ 2004). 
However, Maiolino et al.\ (2004a) found that the UV extinction curve for 
the low-ionization BAL quasar SDSS J1048+4637 at $z=6.2$ is flat at 
wavelengths $\lambda \ga 1700$ \AA~and is rising at $\lambda \la 1700$ 
\AA~, which is quite different from that in the SMC.
Thus, the difference in the extinction curve indicates the different 
origin and evolution of dust at high redshifts ($z > 5$).
Stratta et al.\ (2007) find that the spectral energy distribution of 
afterglow of gamma-ray burst (GRB) 050904 at $z=6.3$ can be reproduced by 
adopting the extinction curve for the $z=6.2$ quasars.
Willott et al.\ (2007) also show that the extinction curve for the 
quasar CFHQS J1509--1749 at $z=6.12$ is similar to that for the $z=6.2$ 
quasar.
These are additional evidence implying that the dust production mechanism 
at $z > 5$ is different from that at $z < 4$.

In this sense, the extinction curve for the $z=6.2$ quasars allows us to 
examine the hypothesis that the formation of dust in the early universe 
is dominated by SNe II and/or PISNe.
Maiolino et al.\ (2004b) have suggested that the extinction curve for the 
$z=6.2$ quasar is in excellent agreement with that derived by using SNe II 
dust models by Todini \& Ferrara (2001).
Models of dust formed in the unmixed SNe II by Nozawa et al.\ (2003) can 
also successfully reproduce the extinction curve observed for the quasar 
at $z=6.2$ by weighting the progenitor mass with the Salpeter IMF 
(Hirashita et al.\ 2005).
However, these studies apply the composition, size distribution, and mass
of dust at the time of its formation and do not take into account
the modification on the physical properties of dust due to the destruction 
in SNRs.
Recently, Bianchi \& Schneider (2007) consider the processing of dust by 
the reverse shock within SNRs and show that the extinction curve by the 
dust surviving the destruction is still consistent with the extinction 
curve for the $z=6.2$ quasar.

Here we present the extinction curve expected at the high-redshift universe 
(Hirashita et al.\ 2008) by using the size distribution and mass abundance 
of each dust species surviving the destruction in SNRs by Nozawa et al.\ 
(2007).
The results of dust evolution within SNRs show that small-sized grains are 
destroyed more efficiently and that the typical size of dust injected from 
SNe to the ISM

\begin{figure}
\plottwo{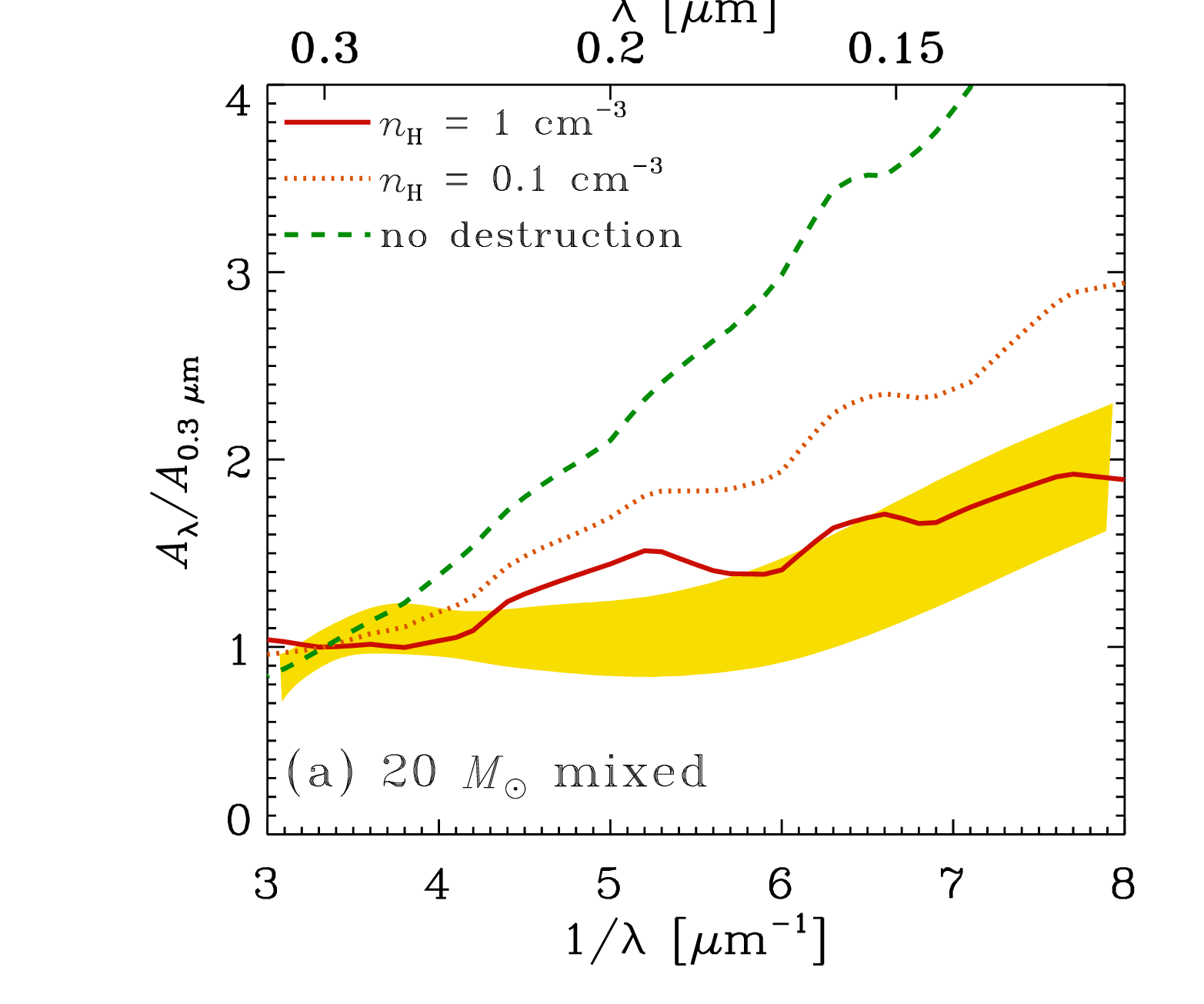}{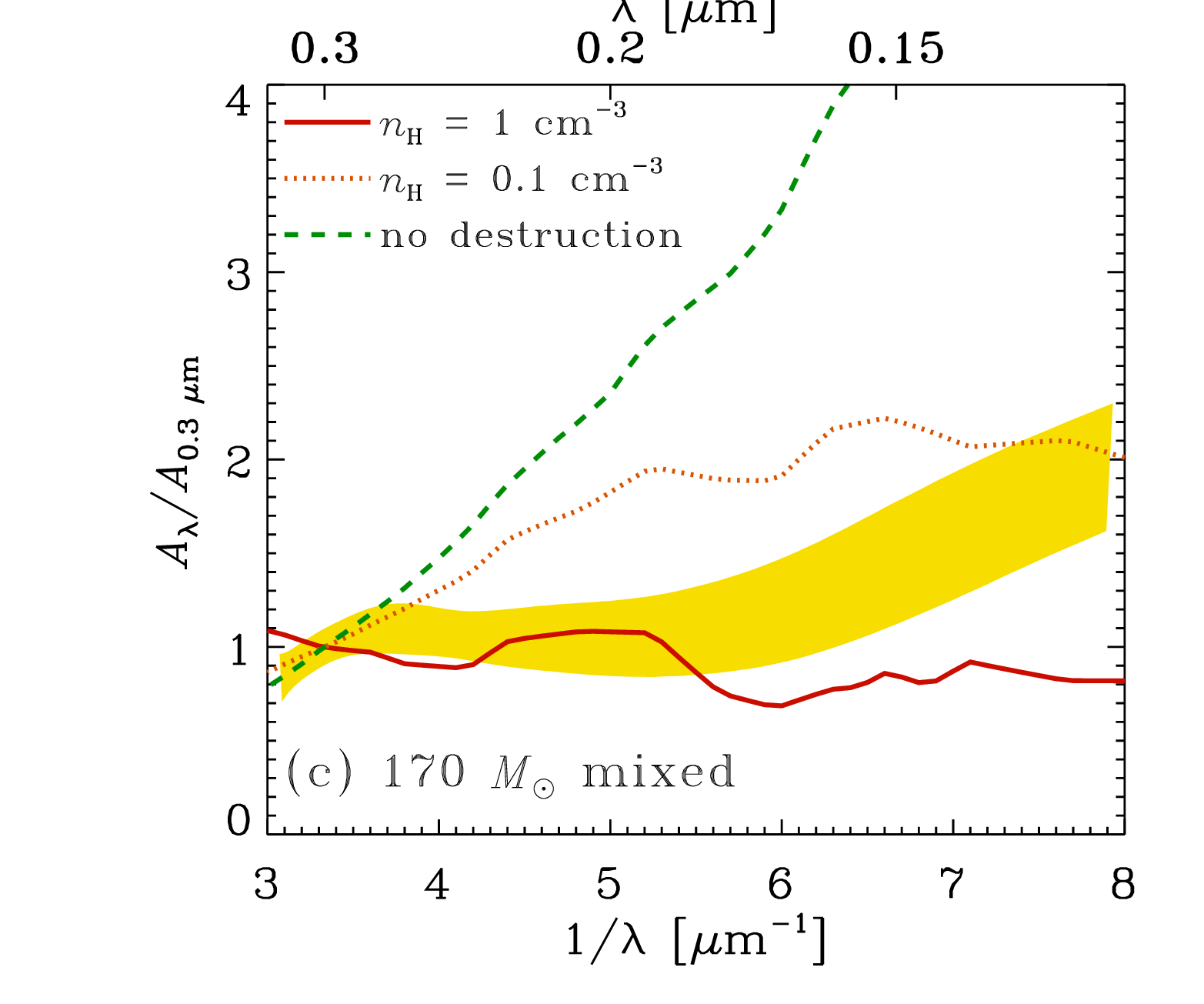}
\plottwo{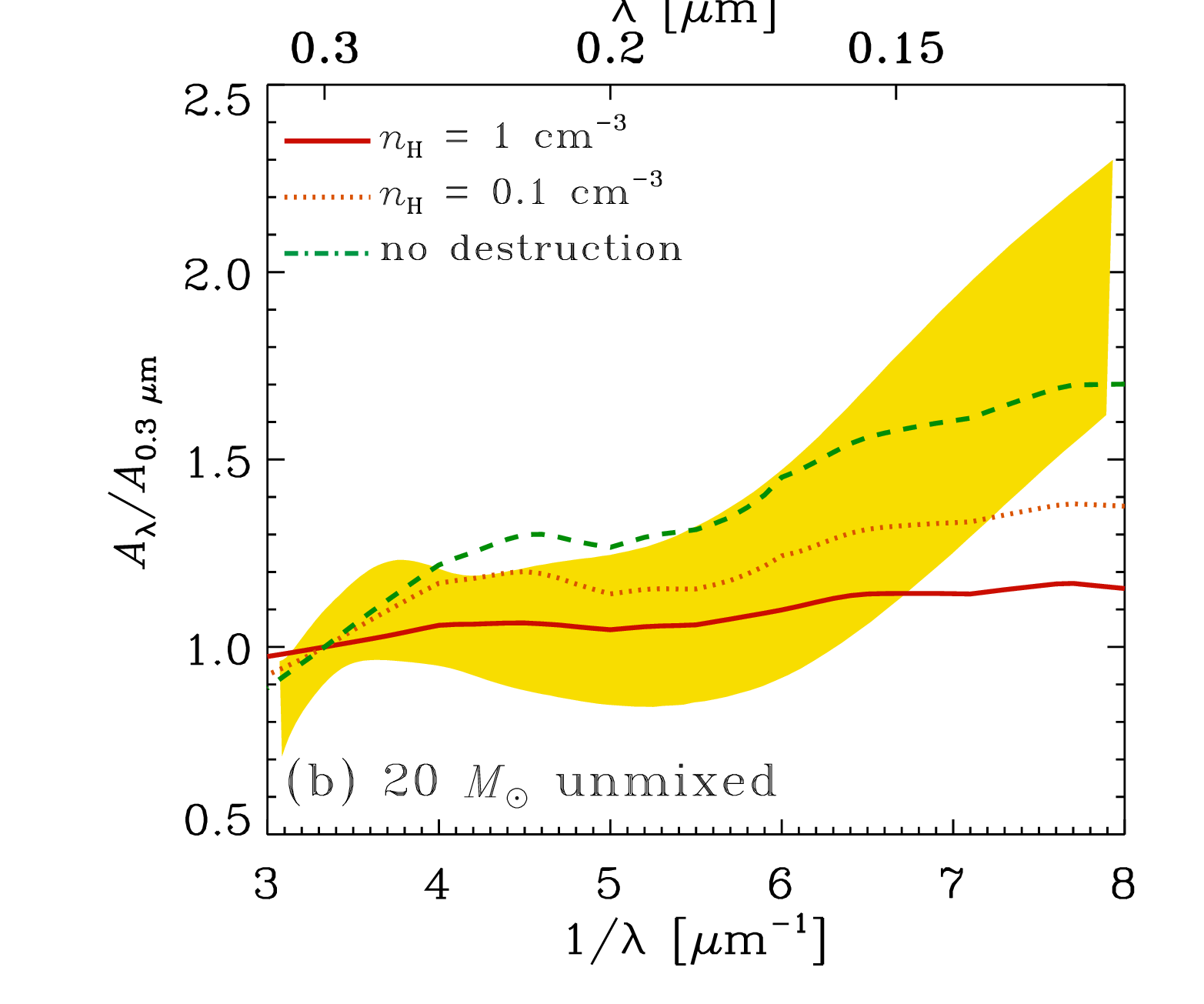}{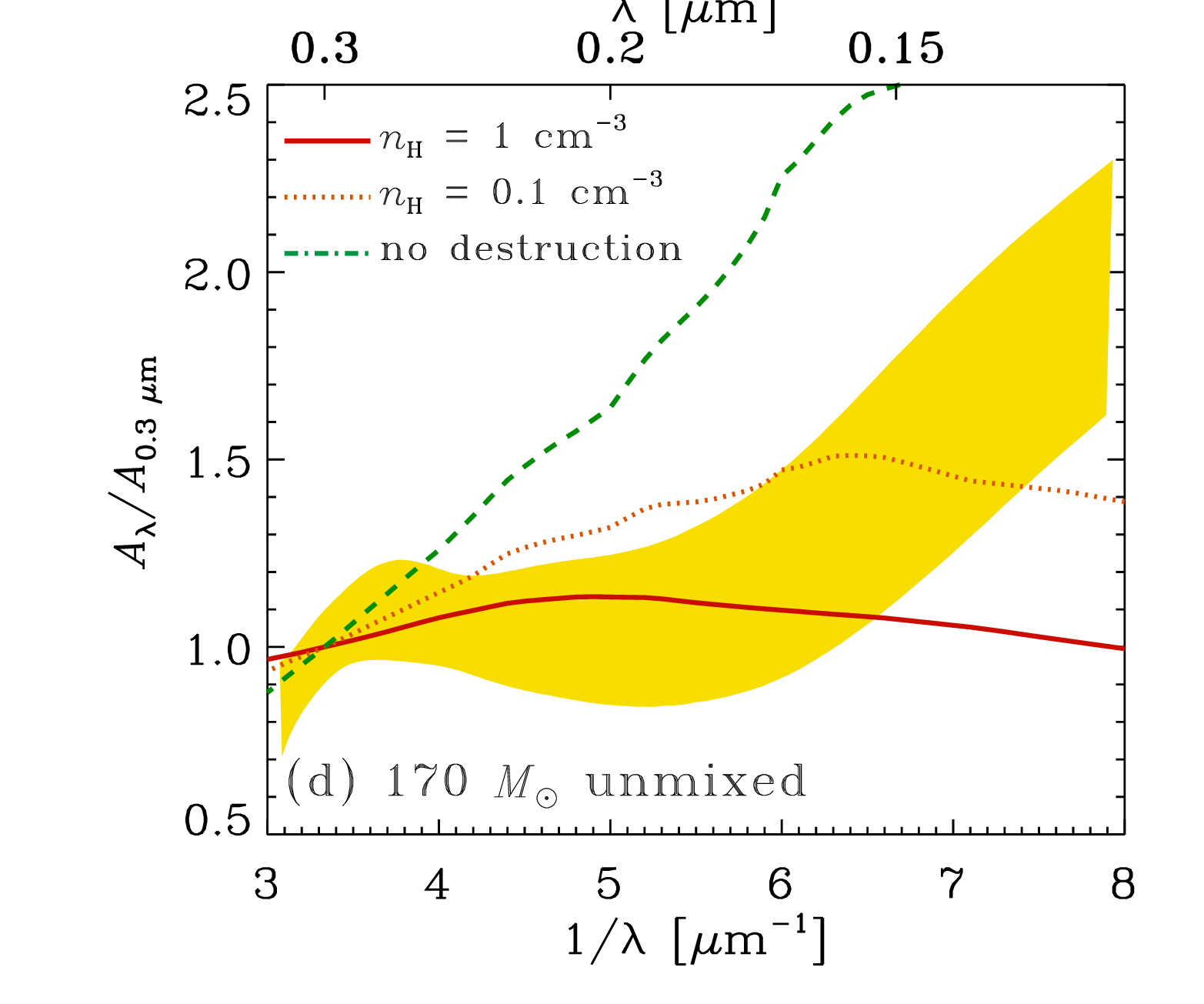}
\caption{
  Dust extinction curves derived for the mixed ejecta with
  ({\it a}) $M_{\rm pr}=20$ $M_\odot$ and
  ({\it b}) $M_{\rm pr}=170$ $M_\odot$, and the unmixed ejecta with
  ({\it c}) $M_{\rm pr}=20$ $M_\odot$ and
  ({\it d}) $M_{\rm pr}=170$ $M_\odot$. 
  The dotted and solid lines indicate the results obtained from the dust
  surviving the destruction in SNRs for the ambient hydrogen number densities 
  of 0.1 and 1 cm$^{-3}$, respectively.
  The case without the reverse shock destruction is also shown by the 
  dashed lines.
  The shaded area in each panel shows the range observed for the quasar 
  SDSS J1048+4637 at $z=6.2$ (Maiolino et al.\ 2004b).
  The extinction is normalized to the value at $\lambda=0.3$ $\mu$m.
}
\end{figure}

\noindent
is biased toward relatively large size ($>$ 0.02 $\mu$m).
Furthermore, for the higher ambient gas density, the efficiency of dust 
destruction is higher, and the average radius of surviving dust shifts 
further toward large radius. 
As a result, the derived extinction curve becomes flatter as the ambient 
gas density increases, regardless of types of SNe and mixing of 
elements within the He core, as shown in Figure 7.
This leads to the conclusion that at high redshift when SNe are the 
possible sources of dust, the dust extinction curves are very flat.
We also compare our results with the extinction curves for the quasars 
at $z=6.2$ (Maiolino et al.\ 2004b).
It can be seen that the derived extinction curves can be also allowed by the 
current observational constraints.

It should be noted here that there are some pieces of evidence supporting 
the flat extinction curve in high-redshift systems. 
Two high-$z$ BAL quasars, SDSS J1044--0125 ($z=5.8$) and J0756+4104 
($z=5.1$) are detected at submillimeter wavelengths (Priddey et al.\ 2003), 
which indicates the presence of dust grains in excess of $10^8$ $M_\odot$.
However, it is reported that these quasars do not show any reddening in 
their UV spectra (Maiolino et al.\ 2004a). 
The absence of reddening and the detection of dust emission seem to be
mutually exclusive, but this contradiction can be resolved by the flat 
extinction curve. 
In addition, even at low redshift, the extinction curves of active 
galactic nuclei (AGNs) are suggested to be flat at UV wavelengths and are 
quite different from those in the diffuse local ISM 
(Maiolino et al.\ 2001a; Gaskwel et al.\ 2004, 2007). 
This also favors the existence of large grains in the circumnuclear 
dense region of AGNs where the coagulation of grains and/or the dominant
depletion of small grains by sublimation are expected to take place
(Maiolino et al.\ 2001b).
However, the extinction curves extracted from SDSS quasar samples 
are similar to that in the SMC and do not show flat behavior in the UV.
This discrepancy is considered to be due to the systematic difference in
the method used in determining extinction laws from composite quasar 
spectra (e.g., Willott 2005).

On the other hand, it has been recognized that the afterglows of GRBs are 
obscured and reddened by dust (Li, A. et al.\ 2008 and references therein) 
and are powerful tools to derive the extinction curves in their host 
galaxies (e.g., Starling et al.\ 2007; Heng et al.\ 2008).
The spectrum of afterglow is given by the power-law formula over the 
broad range from X-rays to near-infrared wavelengths.
Thus, comparing the actually observed spectrum with the power-law 
spectrum extrapolated from the X-ray bands, we can obtain the intrinsic
UV-optical extinction curve.
Using this method, some studies (Stratta et al.\ 2005; Chen et al.\ 2006)
have reported the flat properties for the GRB extinction curves, almost 
independent of wavelengths.
Li, Y. et al.\ (2008) also find that the very flat extinction curves are
identified in the host galaxies of four GRBs with a wide range of redshifts 
($z=0.5$--4). 
These facts indicate that the size distribution of dust is biased toward 
large grains in the environments surrounding the GRBs, which can be 
produced by the coagulational growth of dust in the dense clouds, 
the evaporation of small grains due to the intense X-ray and UV emission 
from the GRBs, or the preferential destruction of small grains by SN 
blast waves.
The latter case is consistent with the destruction of newly formed 
dust in the shocked gas within SNRs considered here.

\section{Concluding remarks}

In this paper, we briefly outline recent advances in theoretical studies 
of the formation and evolution of dust in the early universe, with 
emphasis on the composition, size distribution, and amount of dust ejected 
from primordial SNe.
The investigations of the evolution of newly formed dust in SNRs show 
that the transport and destruction of dust grains within SNRs heavily 
depend on their initial radii.
Small grains are efficiently decelerated by the gas drag and are 
predominantly destroyed by sputtering in the shocked gas.
A fraction of large grains surviving the destruction are finally trapped 
in the dense shell behind the forward shock, and the larger grains are 
injected into the ISM without undergoing significant deceleration and 
erosion.

Dust grains surviving the destruction in SNRs bring a few important 
conclusions.
The dust grains accumulated in the dense SN shell can enrich the 
primordial gas gathered by the forward shock and cause the formation of 
low-mass Population II.5 stars in the dense shell.
Those dust grains can also affect the elemental compositions of the
Population II.5 stars and produce the metal abundance patterns resembling 
those seen in HMP and UMP stars.
In addition, the large grains ejected from SNe prescribe the extinction 
property of the ISM and lead to a flat extinction curve at the
high-redshift universe where dust grains are mainly supplied by 
primordial SNe.
Similar situations appear in the circumnuclear regions of AGNs and the
GRB host galaxies at low redshifts.
Finally, the mass of dust surviving in primordial SNRs reaches up to 
0.1--15 $M_\odot$, which is consistent with the dust yield per SN required 
to explain a large quantity of dust observed in high-redshift systems.
Thus, SNe play the dominant role in the enrichment history of dust 
as dust factories in the early universe.

Recently, Cherchneff \& Lilly (2008) have suggested that primordial SNe
are the first molecule providers to the early universe.
They show that the mixing of hydrogen from the progenitor envelope 
facilitates the formation of molecules in the He core, which acts as a 
bottleneck process against the formation of dust in the SN ejecta.
However, it has been still debated how large the mixing of elements 
extends into the ejecta and whether the mixing is at the knotty level 
or at the atomic level.
We also note that although the mass of dust formed in nearby SNe has been 
considered to be less than $10^{-3}$ $M_\odot$, the estimated mass is 
derived for a limited number of observations by assuming that the ejecta 
is optically thin (see Kozasa et al.\ in this volume, see also 
Sugerman et al. 2006; Miekle et al.\ 2007; Nozawa et al.\ 2008).

In order to reveal the evolution of dust grains in the early universe,
we need further investigations of the fundamental processes of formation 
and destruction of dust expected at high redshift.
Elvis et al.\ (2002) have proposed the possible condensation of a large 
amount of dust in quasar winds, while it is suggested that the formation 
of dust in AGB stars can dominate over the contributions from SNe even at 
$z=5$ (Valiante, R. et al.\ in preparation).
Hence, the formation of dust and its survival in these possible formation 
sites in the early universe should be examined.
In addition, modellings of the spectral energy distribution of high-$z$ 
galaxies (e.g., Takeuchi \& Ishii 2004) as well as nearby young blue compact 
dwarfs (e.g., Takeuchi et al.\ 2003, 2005; Galliano et al.\ 2005) are 
crucial to specify the properties of dust in the primeval universe.
Such sophisticated studies that treat the chemical evolution and the 
radiation transport calculation self-consistently can be powerful probes 
to elucidate the origin and nature of the cosmic dust in the early universe, 
in particular when their results are compared with the future observations 
with {\it Herschel}, {\it JWST}, and {\it ALMA}. 

\acknowledgements %%% Text of acknowledgements runs on after this command.
 This work has been supported in part by World Premier International 
 Research Center Initiative (WPI initiative), MEXT, Japan and by the
 Grant-in-Aid for Scientific Research of the Japan Society for the 
 Promotion of Science (18104003, 19740094, 20340038).

\end{document}